# Theory of Graphene-Insulator-Graphene Tunnel Junctions

S. C. de la Barrera, Qin Gao, and R. M. Feenstra
Dept. of Physics, Carnegie Mellon University, Pittsburgh, Pennsylvania 15213, USA

## Abstract

Graphene-insulator-graphene vertical tunneling structures are discussed from a theoretical perspective. Momentum conservation in such devices leads to highly nonlinear current-voltage characteristics, which with gates on the tunnel junction form potentially useful transistor structures. Two prior theoretical treatments of such devices are discussed; the treatments are shown to be formally equivalent, although some differences in their implementations are identified. The limit of zero momentum conservation in the theory is explicitly considered, with a formula involving the density-of-states of the graphene electrodes recovered in this limit. Various predictions of the theory are compared to experiment.

## I. Introduction

Recently, several research groups have reported theoretical and/or experimental results relating to vertical graphene-insulator-graphene (GIG) tunneling structures. The first such report dealt with coupled electron and hole gases in the two opposing electrodes, predicted to form an exciton condensate that might survive at temperatures as high as room temperature.[1,2] The presence of this condensate leads to an enhanced tunnel current (i.e. since the electrons and holes in opposing electrodes have correlated spatial locations), but for a sufficiently high current the condensate is expected to be quenched. Hence, a very nonlinear relationship of tunnel current to voltage across the device, with negative differential resistance (NDR), is expected. With a gate electrode on the device, a transistor-like operation is achieved in a device termed a *BiSFET*. We are not aware of experimental observation of the BiSFET tunnel characteristic to date, although research on such devices is likely continuing.

Following the BiSFET proposal it was realized by Feenstra et al. that, even in the absence of electron-hole coupling between the graphene electrodes, the *single-particle* tunneling characteristics of GIG devices can be highly nonlinear.[3] The reason for this behavior arises from *momentum conservation* in the device, i.e. the requirement that the lateral components of the wavefunctions for tunneling states in both electrodes have the same (or nearly the same) wavevectors. A theory was developed in which momentum conservation in an actual device was shown to depend on the crystallographic order of the graphene electrodes, which is limited by a finite size tunneling area (grains of the graphene) or through scattering from defects in the graphene or insulator layers.[3] The effective size of ordered regions in the electrodes can be characterized by a *coherence length*, with momentum conservation being more rigorously followed when the coherence length is large.

Experimentally, early results by Britnell et al. from GIG junctions did not display any NDR.[4] Indeed, their theoretical description of such devices employed a theory in which momentum conservation is completely neglected. Similarly, NDR was not seen in early reports from Roy et al. for GIG junctions.[5] However, later results from Britnell et al. *did* reveal NDR in the GIG devices, and a correspondingly more general theory was described in which momentum conservation is included.[6] Related theories have been recently presented by other authors.[7,8,9]

In this work we compare the theoretical description by Britnell et al. for GIG devices[6] to the earlier treatment of Feenstra et al.[3] We find that the two treatments are equivalent, at least in the limit of zero misorientation angle between the graphene electrodes. This equivalence between the two theories, and the possible effects of misorientation, are discussed in the following Section. We also discuss the limit in which momentum conservation is completely neglected,[4] dealing in particular with the problem of how to obtain absolute current magnitudes in that case. In Section III we focus on hexagonal boron nitride (h-BN) barrier materials, describing their complex band structure and hence revealing the energy dependence of the tunneling decay constant. A comparison of the theoretical results with experiment is given in Section IV, and the paper is summarized in Section V.

## II. Theoretical Formalism

In a prior report, Britnell et al. presented experimental data for current-voltage characteristics of a single-gated GIG junction, and interpreted the characteristics using a theory in which momentum conservation is completely neglected.[4] As described in their work, the expression for the current then has the form

$$I \propto \int D_L(E) \, D_R(E) \, T(E) \big[ f_L(E) - f_R(E) \big] dE \tag{1}$$

where $D_L$ and $D_R$ are the densities-of-states for the left- and right-hand electrodes, respectively, $f_L$ and $f_R$ are their Fermi-Dirac occupations factors, and $T(E)$ is a tunneling transmission term. (In this expression the shift in the states and Fermi energies of the two electrodes due to a voltage bias $V$ between them is contained within the $D_L$, $D_R$, $f_L$ and $f_R$ terms, rather than in the energy arguments themselves as done in Ref. [4], so as to be consistent with the formalism presented below).

When momentum conservation (wavevector conservation) for the lateral parts of wavefunctions in the two graphene electrodes is included, then the theory becomes significantly more complex as discussed in Refs. [3] and [6], which employ theories that might appear at first glance to be quite different. We compare those two theories in this section, showing that they are actually equivalent for the situation of zero misorientation angle between the graphene electrodes. We discuss possible effects due to misorientation, and we also identify a few other differences in implementation of the two theories.

In Ref. [3], tunneling between two graphene electrodes is written in the Bardeen formalism,[10,11,12,13] in which the current is given by

$$I = g_V \frac{4\pi e}{\hbar} \sum_{\alpha,\beta} \big| M_{\alpha\beta} \big|^2 \big[ f_L(E_\alpha) - f_R(E_\beta) \big] \delta(E_\alpha - E_\beta). \tag{2}$$

where $g_V$ is the valley degeneracy of graphene, and the summation extends over all states $\alpha, \beta$ of the left- and right-hand electrodes, respectively. The matrix element $M_{\alpha\beta}$ is given by

$$M_{\alpha\beta} = \frac{\hbar^2}{2m} \int dS \left( \Psi_\alpha^* \frac{d\Psi_\beta}{dz} - \Psi_\beta \frac{d\Psi_\alpha^*}{dz} \right) \tag{3}$$

where $m$ is the free electron mass and $\Psi_\alpha(\mathbf{r},z)$ and $\Psi_\beta(\mathbf{r},z)$ are the wavefunctions of the left- and right-hand electrodes (each of those electrodes taken to be connected to a semi-infinite barrier), respectively. For a graphene-insulator-graphene junction, $M_{\alpha\beta}$ is evaluated in Ref. [3] by assuming the wavefunctions to be separable, with exponentially decaying $z$-components and with lateral components that have Bloch form, yielding

$$M_{\alpha\beta} = \frac{\hbar^2 \kappa}{2AmD} e^{-\kappa d} \, g_\omega(\theta_L, \theta_R) \int dS \, e^{i\mathbf{Q}\bullet\mathbf{r}} e^{i(\mathbf{k}_R - \mathbf{k}_L)\bullet\mathbf{r}} \tag{4}$$

where $g_\omega(\theta_L, \theta_R)$ is an expression of order unity that involves the overlap of periodic part of the lateral wavefunctions ($\theta_L$ and $\theta_R$ being the angular orientation of their wavevector relative to the respective Dirac point), $\mathbf{Q}$ is the misorientation vector of the graphene electrodes with corresponding misorientation angle $\omega$, and where $\mathbf{k}_L$ and $\mathbf{k}_R$ are the lateral wavevectors of the states in the left- and right-hand electrodes, relative to their respective Dirac points. All other parameters are defined precisely as in Ref. [3]. Significantly, in Ref. [3] the surface integral of this equation is *restricted in lateral extent*, $L$, for both the $x$ and $y$ directions. This restriction can arise from the lateral extent of the graphene grains in the electrodes, i.e. a "structural coherence length", as proposed in Ref. [3].

Turning to the theory of Ref. [6], the matrix element for the tunneling process is written there as

$$M_{\alpha\beta}^S = \int_V dV \, \Psi_\alpha^*(\mathbf{r},z) V_S(\mathbf{r},z) \, \Psi_\beta(\mathbf{r},z) \tag{5}$$

where the integral extends over all space and $V_S$ is denoted a "scattering potential". In the computations of Ref. [6] this scattering potential is taken to be localized over the region of the tunnel barrier. Although this form appears to be quite different than that of Eq. (3), we demonstrate now that the two methods are equivalent.

Following Ref. [6], Eq. (5) is evaluated as (using notation of the present work)

$$M_{\alpha\beta}^S = \frac{1}{AD} e^{-\kappa d} \, u_{11}^2 \, e^{i(\theta_L - \theta_R + \omega)/2} \, \Xi \int dS \, e^{i\mathbf{Q}\bullet\mathbf{r}} e^{i(\mathbf{k}_R - \mathbf{k}_L)\bullet\mathbf{r}} \tag{6}$$

where we have substituted back into Eq. (S11) of Ref. [6] their expression for $\overline{V}_S^{\parallel}(\mathbf{r})$ from their Eqs. (S8) and (S9). For the purpose of comparing this equation to Eq. (4), we have pulled out from the integrand the periodic part of the Bloch function, i.e. following Ref. [3], to form the $u_{11}^2$ prefactor. Additionally, we have employed the sign convention for misorientation from Ref. [3], so that the signs of $\mathbf{Q}$ and $\omega$ in Eq. (6) are opposite those in Ref. [6]. Now, comparing Eqs. (4) and (6), we note that the expression $g_\omega(\theta_L, \theta_R)$ in Eq. (4) is simply a generalization of the $u_{11}^2 \, e^{i(\theta_L - \theta_R + \omega)/2}$ terms in Eq. (6) (as shown in the latter part of the derivation in Ref. [6]). With that, we find that Eqs. (4) and (6) produce identical results so long as we take $\Xi = \hbar^2 \kappa / 2m$. In terms of the scattering potential of Eq. (6), assumed as in Ref. [6] to be separable with $V_S(\mathbf{r}, z) = V_S(z) V_S^{\parallel}(\mathbf{r})$, this value of $\Xi$ corresponds to $V_S(z) = \hbar^2 \kappa / 2md$ for the case of $V_S(z)$ assumed to be constant over the barrier region. Thus, if Eq. (6) is used for computing the tunnel current, then this specific magnitude of $V_S$ must be employed (or, for a varying $V_S(z)$ across the barrier, some generalization of this magnitude could be obtained, again through the use of Eqs. (3) and (4)). With this specific value, the tunneling formalism of Ref. [6] is then seen to be identical to that of Ref. [3].

It should be noted that our comparison of Eqs. (3) and (4) with (5) and (6) is made on the assumption that the latter equations are being used to compute the total (or primary) tunnel current. Alternatively, if some secondary source of scattering in the system is assumed, then Eq. (5) can be applied more directly, with some arbitrary (assumed) value of the scattering potential. This distinction is emphasized by Duke,[11] where he refers to the primary contribution as the "elastic coherent" one, computed using a matrix element like that of Eq. (3), and with any secondary contribution computed according to a matrix element like that of Eq. (5) (see, e.g. Eq. (18.38) of Ref. [11]). In such a computation, however, the secondary current would be *summed* together with the primary one. Use of such a summation is not discussed by Britnell et al.,[6] and so we interpret their equation as indeed being intended for expressing the total tunnel current.

Despite the equivalence in the formalisms of Refs. [3] and [6], there are a number of differences in the implementation of their theories for producing numerical results. First, in Ref. [3] a specific model for the tunnel barrier was not considered beyond what would be appropriate for a vacuum barrier (i.e. isotropic band with effective mass of unity). In this respect the treatment of Ref. [6] for a specific barrier material (such as h-BN) is significantly better. In Section III we extend that sort of treatment, providing theoretical results for the energy dependence of the tunneling decay constant.

A second difference in implementation has to do with the specific means of evaluating the matrix elements. Consider the surface integral in Eqs. (4) and (6), normalized to the area $A$ of the junction,

$$\frac{1}{A} \int dS \; e^{i\mathbf{Q} \bullet \mathbf{r}} \; e^{i \Delta \mathbf{k} \bullet \mathbf{r}} \tag{7}$$

where $\Delta \mathbf{k} \equiv \mathbf{k}_R - \mathbf{k}_L$. In Ref. [3], this term is evaluated over a finite range, $-L/2$ to $+L/2$ for both $x$ and $y$ directions, which for zero misorientation leads to

$$\mathrm{sinc}\left(\frac{L \Delta k_x}{2}\right) \mathrm{sinc}\left(\frac{L \Delta k_y}{2}\right). \tag{8}$$

For the case of Ref. [6], this part of the matrix element is captured in their $\overline{V}_S^{\parallel}(\mathbf{q})$ term with $\mathbf{q} \equiv \Delta \mathbf{k}$, which similarly restricts the region over which the tunneling occurs in a laterally coherent manner. A quantity analogous to that in Eqs. (7) or (8) would be $\overline{V}_S^{\parallel}(\mathbf{q}) / A$, which in Ref. [6] is modeled as

$$\frac{1}{A(q_c^2 + q^2)} \tag{9}$$

where $q_c$ is some cut-off wavelength. If we compare the tunnel currents obtained using Eqs. (8) and (9), we find fairly good agreement in the dependence of the current on the parameters $L$ and $q_c$, so long as we take $L = 2\pi q_c^{-1}$. However, regarding the absolute magnitude of the current, we find poor agreement between that obtained from Eqs. (8) and (9), even with the use of $\Xi = \hbar^2 \kappa / 2m$. This problem arises from the specific dependence of Eq. (9) on the area $A$ of the device, which produces an incorrect dependence of the current on $A$ (it should be noted that Eq. (9) was presented in Ref. [6] primarily as a proportionality, i.e. without focus on the absolute magnitude of the term). However, if we modify the form of Eq. (9) somewhat we *can* obtain current that scales properly with $A$. In particular, we use

$$\frac{2\pi q_c^2}{\sqrt{A}\,(q_c^2 + q^2)^{3/2}} = \frac{1}{[1 + (q/q_c)^2]^{3/2}} \tag{10}$$

Equation (10) produces very similar results as Eq. (8) in terms of both the parameter-dependence and the absolute magnitude of the current, still considering zero misorientation.

The equivalence between the two theoretical treatments is illustrated in Fig. 1(a) and (b), showing a side-by-side comparison of tunneling currents computed using Eqs. (8) and (10), respectively, with related parameters $L = 2\pi q_c^{-1}$ and $\Xi = \hbar^2 \kappa / 2m$. Although the results are qualitatively similar, we consider Eq. (10) to be slightly preferable compared to Eq. (8) for evaluation of the current, since the latter employs sharp cut-offs for a single $L$-value in the $x$ and $y$ directions, which produce small oscillations in the current-voltage characteristic above the main resonant peak.[3] These oscillations are not present when Eq. (10) is employed, since that equation is applicable to a *distribution* of $L$-values, as is likely more appropriate for physical device. We show this equivalence explicitly in Fig. 1(c), which is obtained by computing the total current density for a polycrystalline device with a log-normal distribution[14] of grain sizes (i.e. a distribution of coherence lengths). Including such a distribution of grains in a single device averages out secondary oscillations due to grain size effects but preserves the resonant peak structure and yields a tunneling characteristic similar to that of Eq. (10), shown in Fig. 1(b). Compared to a computation involving a distribution of grains and multiple calculations with Eq. (8), a straightforward computation using Eq. (10) appears to capture the relevant physics of a macroscopic device in a more compact form, and thus we use Eq. (10) in all subsequent calculations.

We conclude that the theories of Refs. [3] and [6], employing Eqs. (8) or (9), respectively, are actually modeling the same aspect of the tunneling process, namely, a restriction in the lateral extent over which the wavefunctions maintain their coherence. In Ref. [3] this was described in terms of a grain size in the graphene. In Ref. [6] this was described in terms of the "scattering potential" of Eq. (5), with specific form given by Eq. (9) (or with a small modification to that, as in Eq. 10). Again, the effect of this "scattering potential" is to restrict the lateral area over which coherent tunneling occurs. However, Ref. [6] it is argued that this restriction is not due to limited grain sizes in their devices, but rather, arises from other scattering mechanisms in the system.

Another significant difference between the theories of Refs. [3] and [6] is in the manner in which they deal with angular misorientation between the lattices of the graphene electrodes. For Ref. [6], it is assumed that there is no dependence on misorientation, with Eq. (9) being used in the computations where $\mathbf{q} = \Delta\mathbf{k}$ as defined following Eq. (7). That is to say, the factor $e^{i\mathbf{Q}\bullet\mathbf{r}}$ in Eq. (7) is incorporated in their definition of a modified scattering potential $\overline{V}_S^{\parallel}(\mathbf{r})$, so that the Fourier transform of that quantity, $\overline{V}_S^{\parallel}(\mathbf{q})$, can be modeled directly by Eq. (9) without any further explicit occurrence of the $e^{i\mathbf{Q}\bullet\mathbf{r}}$ term. This treatment thus makes a specific assumption about the scattering mechanism (although the specific physical mechanism is not identified).

In contrast, in Ref. [3] the misorientation is fully included, employing $\mathbf{Q} + \Delta\mathbf{k}$ in the argument of the combined exponentials of Eq. (7) where $\mathbf{Q}$ is the misorientation vector. Similarly, writing Eq. (8) with inclusion of misorientation we would have $Q_x + \Delta k_x$ and $Q_y + \Delta k_y$ in the arguments of the sinc functions, as evaluated in Ref. [3], rather than just $\Delta k_x$ and $\Delta k_y$. For the present work in which we use the more general form given by Eq. (10), we also evaluate that with $|\mathbf{Q} + \mathbf{q}|$ in the argument rather than just $q$. This procedure is followed for all subsequent computational results in this work, so that using Eqs. (4) and (10) our matrix elements are computed as

$$M_{\alpha\beta} = \frac{\hbar^2 \kappa}{2mD} e^{-\kappa d} \frac{g_\omega(\theta_L, \theta_R)}{[1 + (|\mathbf{Q} + \mathbf{q}|/q_c)^2]^{3/2}} \tag{11}$$

with $\mathbf{q} = \Delta\mathbf{k}$. The current is then given by Eq. (2).

Regarding the role of misorientation (as determined by $\mathbf{Q}$), we find that this is a large effect, consistent with the results of Ref. [3]. In Ref. [6], misorientation is handled by absorbing the $e^{i\mathbf{Q}\bullet\mathbf{r}}$ from Eq. (7) into their definition of the scattering potential $\overline{V}_S^{\parallel}(\mathbf{q})$. We do not agree with their argument that the resulting current-voltage relationship will not show a significant dependence on misorientation. Certainly for small $L$ (large $q_c$) misorientation is not so important, but we feel that in general the misorientation *will* play a large role in determining the current-voltage characteristic. We thus feel that it is best to leave this issue as an open question for the moment, hopefully to be addressed experimentally in future work.

Summarizing this comparison of the theories of Refs. [3] and [6], we find the following: (i) The two theories are formally equivalent, although we find that the $\Xi$ parameter in the latter theory must have a value of $\hbar^2 \kappa / 2m$ (and also $L$ is related to $q_c$ by $L = 2\pi q_c^{-1}$). (ii) The scattering term in the latter theory is slightly modified here, as in Eq. (10). With that revision, numerical results from the two theories are in good agreement for the case of zero misorientation. (iii) For nonzero misorientation, we believe that the former theory provides the correct form for the tunneling current at least when finite grain sizes limit the lateral coherence of the tunneling. For other scattering mechanisms perhaps misorientation is not so important, as assumed in Ref. [6], although specific identification of such a mechanism remains to be done. Further work, both experimental and theoretical, is likely needed to evaluate the role of electrode misorientation in the tunneling.

We briefly comment on one additional aspect of the tunneling formalism, namely, the use of the density-of-states formula of Eq. (1) for computing tunneling current.[4] This formula is commonly used in tunneling computations, although obtaining an absolute magnitude of the current is problematic with this approach since it is not obvious what the appropriate pre-factors in front of the integral should be. Of course, with the full theory of Eq. (2), we can obtain a current with well-defined magnitude. Also, in the limit of $L \to 0$ of that theory, it is easily shown that we recover Eq. (1). However, when we compute currents in that limit, i.e. for smaller and smaller $L$ values, then the currents that we obtain (actually they are current densities, since the computation is for a specific $L^2$ area) become unphysically small. The question we must address is, what is the fundamental source of this decrease in current density for $L \to 0$, and can we somehow produce a current density whose magnitude is physically meaningful even in this limit.

The origin of the unphysical $L \to 0$ limit of the full theory of Eq. (2), when evaluated together with Eq. (4) or (6) and Eq. (7) or (10), arises from our assumption of limiting the area over which the surface integral in Eq. (2) is performed. For very small $L$ values, we then encounter a situation in which the tunneling is restricted to a small area of one electrode over to the same small area of the opposite electrode. This restriction is invalid since we are ignoring the tunneling to *neighboring* areas in the opposing electrodes. That is, we must consider *spreading* (dispersion) of these states as they extend across the barrier. To properly deal with this situation, we construct states on each electrode that are restricted to an area $L$, hence with wavefunctions proportional to $[H(x+L/2)-H(x-L/2)][H(y+L/2)-H(y-L/2)]\exp(i\mathbf{k}\bullet\mathbf{r})$ where $H(x)$ is a Heaviside step function. We Fourier transform these wavefunctions in order to deduce their dispersion in the barrier, with each Fourier component extending into the barrier with an exponential decay constant $\kappa' = \sqrt{\kappa^2 + \eta^2}$ (assuming equal effective masses in the lateral and perpendicular directions) where $\eta \equiv |\boldsymbol{\eta}|$ denotes the lateral wavevector variable in the Fourier transform. On each electrode the total wavefunction is written as a summation of such states, localized on adjoining areas. We then work through the Bardeen formalism. For a given state restricted to an area $A = L^2$ of the left-hand electrode, we can evaluate contributions to the matrix element Eq. (3) from the overlap of that state with states from all areas of the right-hand electrode. To illustrate our result, we compare it to the surface integral in Eq. (7), for the case of zero misorientation and where we include a $\kappa e^{-\kappa d}$ term in that integrand (i.e. from the prefactor of Eq. 4). Whereas Eq. (8) was obtained by using an *ad hoc* restriction of this surface integral over the area $A$, we now have a more rigorous treatment using our constructed wavefunctions. The term analogous to Eq. (7) then becomes

$$\frac{1}{A}\int dS\, \kappa e^{-\kappa d}\, e^{i\Delta\mathbf{k}\bullet\mathbf{r}} \to \frac{A}{(2\pi)^2}\sum_{m,n}\int_{-\infty}^{\infty}d\eta_x\int_{-\infty}^{\infty}d\eta_y\, \kappa' e^{-\kappa' d}\, \text{sinc}\left(\frac{(\eta_x - k_{R,x})L}{2}\right)\text{sinc}\left(\frac{(\eta_y - k_{R,y})L}{2}\right)$$

$$\text{sinc}\left(\frac{(\eta_x - k_{L,x})L}{2}\right)\text{sinc}\left(\frac{(\eta_y - k_{L,y})L}{2}\right)e^{i[(\eta_x - k_{R,x})mL + (\eta_y - k_{R,y})nL]} \quad (12)$$

where $m$ and $n$ label areas of the right-hand electrode, both extending over $0, \pm 1, \pm 2, \ldots$.

The $m = n = 0$ term of the summation on the right-hand side of Eq. (12) dominates for large $L$, and in that case the expression on the left-hand side of the equation (evaluated as in Eq. 8) is recovered. The additional terms in that sum are negligible for $L > 10$ nm, but they make important contributions for smaller $L$ values. Performing the complete summation for small $L$ values becomes computationally demanding. However, we find for the parameters of our simulations described in Section IV (1.34-nm-wide tunnel barrier with tunneling decay constant 6 nm$^{-1}$), the results of the full summation for $L \to 0$ matches well to the result of including only the $m = n = 0$ term but with the *fixed* value of $L = 1.4$ nm. Therefore, to incorporate an absolute scale of current densities on computations employing Eq. (1), we can simply adjust the magnitude of the results so that they match that of a computation employing Eq. (4) together with Eq. (10) using $L = 1.4$ nm. (We note that the voltage-dependence of the computation using Eqs. (4) and (10) with $L = 1.4$ nm is very close to that obtained with Eq. (1), so in principle we could simply use the former to report the results. Nevertheless it is desirable to use the latter for computations in which no trace of momentum conservation is evident in the experimental data, while at the same time including an estimate of the absolute magnitude for those current densities. We achieve that goal by matching the magnitudes of the two computational results). Of course, this same procedure would be necessary (and would yield similar results) if employing the theory of Ref. [6], i.e. Eq. (6) together with Eq. (9) or (10), for very large $q_c$ values.

### III. Hexagonal Boron Nitride Tunneling Barrier

In the work of Britnell et al., some specific details of a tunneling barrier consisting of hexagonal boron nitride (h-BN) were described.[4] We extend those considerations here by considering the results of explicit computations of the h-BN band structure. In Fig. 2(a) we display the band structure of h-BN along various high symmetry directions, computed using density-functional theory with the Vienna Ab Initio Simulation Package (VASP). We use the Perdew-Burke-Ernzerhof (PBE)[15] parametrization of the generalized gradient approximation (GGA) for the electron exchange correlation potential. We use projector augmented wave potentials[16,17] with a fixed energy cutoff of 400 eV (the default for N). The cell is fixed with experimental lattice constants in the calculations. The zero of energy in Fig. 2(a) is chosen to be coincident with the top of the valence band (VB); a band gap of 4.21 eV separating the VB and the conduction band (CB) is found in our density-functional computation, significantly less than the experimental value of 6.0 eV,[18] with this error occurring due to the well-known limitations of density-functional theory.

For tunneling, we require the band structure for *complex* values of the wavevector $\mathbf{k}$, as discussed in Ref. [4] by employing simple models for the band structure for real $\mathbf{k}$ values and then analytically continuing those to imaginary $\mathbf{k}$ values. The general behavior of such analytic continuation can be deduced from inspection of complex band structures for other materials,[19,20] namely, that the curvature of bands reverses sign when crossing from real to imaginary $\mathbf{k}$ across some critical point in the band structure, but with the *magnitude* of curvature (effective mass) being maintained. If the bands with real **k** approach a critical point with a nonzero slope (as occurs when the Fourier component of the potential for that particular **k** value is zero), then no continuation of the band into imaginary **k** values occurs. Additionally, considering whether or not a band with imaginary **k** value will serve to *connect* bands with purely real **k** (i.e. connecting the VB and CB of h-BN), then the respective states for the two bands at the critical point must have nonzero overlap,[19] i.e. $\langle \psi_i | \psi_j \rangle \neq 0$ for a band with imaginary **k** connecting states $\psi_i$ and $\psi_j$.

To explicitly obtain the complex band structure for h-BN, we employ a tight-binding model with parameter values adjusted such that the bands approximately match those of the density-functional computation (except for the band gap, where the experimental value of 6.0 eV is matched).[18] Results are shown in Fig. 2(b), where we have used a model with only $p_z$-states on the B and N atoms as basis functions (on-site energies of 6.0 and $-0.9$ eV, respectively), and assuming both in-plane and out-of-plane nearest-neighbor B-N interactions (hopping energies of $-1.6$ and 0.6 eV, respectively) as well as a second-nearest-neighbor in-plane N-N interactions ($-0.3$ eV). Additionally, non-orthogonality between both in-plane and out-of-plane nearest-neighbor B-N $p_z$-orbitals is included (overlap matrix elements of 0.05 and 0.03, respectively). The method of solution for this problem with the non-orthogonal basis is described, e.g., in Ref. [21]. Our tight-binding results are similar to those of Robertson.[22]

Comparing Figs. 2(a) and (b), we see that the states derived from the $p_z$-orbitals are quite clearly apparent in the density-functional results. Some mixing occurs with the other, $sp^2$-derived states of the system, with the mixing being strongest in the conduction band. However, for our purposes of evaluating the tunneling of states with large in-plane momentum (near the K or M points), then we note in particular that along the KH and ML directions the tight-binding description of the system using only the $p_z$-orbitals works quite well since the $sp^2$-derived states are separated from the VB and CB edges by about 5 eV. In terms of quantitative agreement between the tight-binding and density-functional results, the former overestimates the band widths for the $p_z$-states along KH (these bands are very flat in the density-functional results) and it underestimates the band widths for the $\Gamma A$ direction. Along ML, the band widths for the tight-binding and density-functional results are reasonably close, within 15%, and those values are also in fairly good agreement with many-body computational results.[23]

From the tight-binding model we can obtain the complex band structure, shown in Fig. 2(c). Those plots are displayed with the same format as Ref. [19]. For example, on the far right-hand side of the plot along the KH direction, in the panel with varying $\mathrm{Im}(k_z)$, there is a loop connecting the VB maximum and CB minimum. This loop is shown by a solid line, indicating that the $\mathrm{Re}(k_z)$ value for these states is *constant*, i.e. it has a value corresponding to the H point, $\mathrm{Re}(k_z) = \pi / c$, where $c$ is the lattice constant of 6.66 Å. For states with energies within the band gap having lateral wavevector corresponding to the K-point, then they will decay in the h-BN with decay constant of $\kappa \equiv \mathrm{Im}(k_z)$ according to the values shown by this loop shown on the far right of Fig. 2(c). The wavefunctions of these states will, at the same time, have a spatial oscillation given by $\mathrm{Re}(k_z) = \pi / c$. This result of a combined exponential decay plus oscillation is a basic feature of the h-BN eigenstates in the [0001] direction through the material (states that have exponential decay *without* any oscillation are not eigenstates of the system).

Turning to the ML and ΓA directions shown in Fig. 2(c), the situation is more complicated. The dashed lines seen there in the $\mathrm{Re}(k_z)$ and $\mathrm{Im}(k_z)$ panels indicate eigenstates for which *both* $\mathrm{Re}(k_z)$ and $\mathrm{Im}(k_z)$ are varying as a function of energy.[19] Focusing on the results in the ML direction, we find a maximum value of $\kappa \equiv \mathrm{Im}(k_z) = 5.2$ nm$^{-1}$ for the (dashed) loop connecting the VB and CB states, at an energy in the middle of the band gap. For the KH direction, at midgap we find a $\kappa$ value of 4.6 nm$^{-1}$, although as discussed above our tight-binding KH bands show too much dispersion; flatter bands are expected to considerably increase this estimated $\kappa$ value. Averaging over angles, we estimate a midgap $\kappa$ value of ≥5.0 nm$^{-1}$. An improved treatment of the complex band structure will provide a better estimate of this value, as well as possibly producing a significant dependence of $\kappa$ on the angle between the graphene and h-BN lattice.

Regarding the energy dependence of $\kappa$, we have found in Fig. 2(c) that we have loops connecting the VB and CB. In the *absence* of a loop, it is usual to model the energy dependence as being parabolic with the energy $\Delta E$ to a band edge, $\kappa = \sqrt{2m^* \Delta E} / \hbar$ with some effective mass $m^*$.[3,6] Now, including the loop, we use this same formula for $\kappa$ but with an interpolation formula for an *effective* barrier height $\Delta E$,

$$\Delta E = \frac{(E_C - E)(E - E_V)}{(E_C - E_V)} \tag{13}$$

where $E_V$ is the energy of the VB maximum, $E_C$ is the energy of the CB minimum, and $E$ is the energy of a state within the band gap. For a midgap $\kappa$ value of $\kappa_0$, the effective mass is given by $m^* = 2\hbar^2 \kappa_0^2 /(E_C - E_V)$.

An experimental value for the tunneling decay constant is available from a prior work of Britnell et al.;[24] computing the slope of their measured tunneling resistance (on a logarithmic scale) as a function of number of BN layers, we find a decay constant of 6.0 nm$^{-1}$. The relationship of this value to the midgap $\kappa_0$ value depends on the offset between the boron nitride VB and the Dirac point of the graphene. Britnell et al. have used an offset of 1.5 eV (i.e. one quarter of the way up the band gap),[4,25] and as discussed in the following Section they have argued that this value accounts for an observed asymmetry in their device characteristics. As also discussed there, from comparison of theory to experiment for other devices we derive an offset value closer to the middle of the band gap.[26] In any case, in order to be definite as to our choice of decay constant to use in our simulations, we take the experimental value of 6.0 nm$^{-1}$ and we assign that to the midgap $\kappa_0$ value. This experimental value is slightly greater than that derived from the tight-binding model discussed above, but still in reasonable agreement considering the uncertainties of the theory. The $\kappa_0$ value of 6.0 nm$^{-1}$ corresponds to an effective mass of 0.9 times the free-electron mass.

## IV. Comparison to Experiment

In this Section we display various simulated results for the GIG current-voltage characteristics, selected to provide comparison to experimental results published elsewhere.[6] The device structures that we consider include either a single gate on the bottom of the device, or both top and bottom gates sandwiching the main GIG structure. Voltages on the gates are denoted $V_{BG}$ and $V_{TG}$ for the bottom and top, respectively. We denote the two graphene electrodes as the source and drain, with the drain being the electrode closest to the top gate and the source closest to the bottom gate. Voltages on the electrodes are denoted $V_S$ and $V_D$ for the source and drain, respectively. We consider the current into the drain, $I_D$, as a function of $V_{DS} \equiv V_D - V_S$. Gate voltages are similarly referenced to the source voltage. In all subsequent simulations we use the two-sided tunneling barrier described by Eq. (13). We calculate carrier densities in the graphene electrodes using the temperature-dependent integrals given in Eq. (27) of Ref. [3], in contrast to the zero-temperature approximation employed in the computational results of our previous work.[3,26,27]

We first consider results obtained on devices that do *not* display NDR in their characteristics, presumably due to a relatively small coherence length for the tunneling. In Fig. 3 we display computed characteristics for a device whose structure (tunneling barrier thickness and gate dielectric thickness) is identical to that employed by Britnell et al., Fig. 4 of Ref. [28]. This device did not display any NDR, and thus we simulate the characteristics with a coherence length of $L \to 0$ (that is, employing Eq. (1) and correcting the magnitude of the current according to the discussion following Eq. 12). Our computed curve for zero gate voltage is essentially identical with that of Britnell et al., and in Fig. 3 we display curves for various other gate voltages as well. Regarding the dependence of the zero-bias conductance on gate voltage, Britnell et al. observed distinct asymmetry with respect to the polarity of the gate voltage, and from that they concluded that the valence band offset between the h-BN and graphene was approximately 1.5 eV. Our computation of this gate voltage dependence, shown in the inset of Fig. 3, agrees qualitatively with those of Ref. [28], though our simulation uses the modified form of the energy dependence of $\kappa$ as given by Eq. (13) and the temperature-dependent carrier densities mentioned in the previous paragraph (whereas Britnell et al. appear to use the zero-temperature form of the carrier densities).

In Fig. 4 we display computed characteristics for a device whose structure is identical to that employed by Roy et al., Fig. 5 of Ref. [26]. Again, this device did not display any NDR and we simulate it in the limit of $L \to 0$. We see a sloping feature in the curves near $V_{DS} \approx 0.25$ V, which corresponds to the Fermi level in the top graphene electrode passing through the vanishing density-of-states at the Dirac point. There are generally two such features in a given current-voltage curve—one for each electrode as the Fermi level passes through the Dirac point. The sloping feature described here is due to the same phenomena as the plateau feature described in Ref. [26], though it is less distinct due to the broadening effect of finite temperature. Our computed zero-bias conductance vs. gate voltage curve is shown in the inset of Fig. 4. In this case, we find agreement between experiment and theory for a valence band offset of 3 eV (solid curve), i.e. with the graphene Dirac point closer to the middle of the h-BN band gap. If we use an offset of 1.5 eV as in Fig. 3, we obtain the curve shown by the dashed line in the inset of Fig. 4, which does not compare well to the experiment. This difference between the offsets obtained for the devices of Figs. 3 and 4 is not understood at present, although measurements for additional device structures will hopefully serve to clarify this situation.

Let us now turn to devices that *do* display NDR in their characteristics, indicative of larger coherence lengths. Figure 5 shows simulated results for the device structure of Britnell et al., Fig. 1 of Ref. [6]. This device has essentially the same structure as the device in Ref. [28], yet exhibits clear NDR for a similar range of gate voltages. Our simulations of this device in Fig. 5(a) – (c) use a lateral coherence length of 75 nm and a valence band offset of 1.5 eV (although the results were not sensitive to the precise value of the offset). Fig. 5(a) shows the result of our theory for zero misorientation angle between the graphene and the h-BN lattices. The resonant peak behavior is in good agreement with the experiment; they are also very close to the simulation results of Britnell et al., since, as argued in Section II, our theory and their theory are essentially equivalent for the case of zero misorientation angle. Shown in Fig. 5(b) and (c) are results for other possible values of the misorientation. We see that a relatively small misorientation angle changes the tunneling current characteristics significantly, shifting the resonant peaks out to larger bias voltages, as well as flattening out the currents at low bias. This shift in voltage is caused by the addition of the misorientation vector **Q** to the momentum conservation condition, which pushes the resonance condition out to higher voltages.[3] For certain doping situations (with nonzero misorientation), there is one positive and one negative peak in the tunneling characteristic due to the symmetry between the conduction and valence bands in graphene near the Dirac point. We see such a peak develop for both signs of $V_{DS}$ in Fig. 5(c) (with a misorientation angle of 1.0°) over a wide range of gate voltages. Whether or not this effect of misorientation can be observed in an experimental device remains an open question.

The overall scale of our computed currents shown in Fig. 5 is significantly larger than what has been observed experimentally, despite the fact that the simulation parameters are partially derived from the measured value of $\kappa_0 \approx 6.0$ nm$^{-1}$, as discussed in Section III. In addition, neither our theory, nor the theory of Britnell et al. can account for the apparent linear background current observed in the devices with NDR, as seen in Fig. 1 of Ref. [6]. One way to produce a linear background current in the simulation is to average over all angles, as in Fig. 7 of Ref [3], however a range of misorientation angles does not appear to be consistent with this experimental device. Further work is needed in order to resolve these discrepancies between the theoretical and the experimental current-voltage characteristics.

## V. Summary

In summary, we have investigated a number of theoretical issues relating to GIG tunnel junctions. Conservation of lateral momentum in such devices leads to nonlinear current-voltage characteristics of the junction, with a resonant peak occurring when the Dirac points of the graphene electrodes are aligned.[3] Addition of gate electrode(s) can then produce transistor-type behavior of the devices.[27] Theories describing the characteristics of the devices have been previously presented in Refs. [3] and [6]. Despite the seemingly different derivations used for the two theories, we have demonstrated here that they are actually equivalent. In both cases, a limitation of the lateral coherence length leads to broadening of the resonant peak. However, an important distinction between the two theories is in the manner in which misorientation of the graphene electrodes is treated; it is fully included within the theory of Ref. [3] in which the limitation of lateral momentum is assumed to arise through some limited area over which the tunneling occurs, whereas it has no significant effect in the theory of Ref. [6] since the misorientation is folded into the "scattering potential" of the problem. The recent theoretical work of Brey[29] fully includes misorientation effects in the same manner as in Ref. [3].

Experimental results for GIG junctions have been reported,[4,6,26] some of which apparently display little or no momentum conservation, i.e. no resonant peak, and others of which *do* display a resonant peak. In the former case the results can be simulated with a simple formula involving only the density-of-states of the electrodes;[4] we have used that formula here for simulating recent experimental data[26] and we have also argued how the absolute magnitude of the current in this type of computation can be determined. For data in which a resonant peak *is* observed, we investigate the possible effect of electrode misorientation on the results. At least for the data reported thus far, we find that the best comparison with simulation occurs for zero misorientation angle, a conclusion which is apparently consistent with the theory of Ref. [6] since it explicitly neglects the role of misorientation. The reason for this lack of dependence on misorientation angle is not clear at present.

Separately, we have investigated the complex band structure of the h-BN tunneling barrier material. The values of the tunnel decay constant $\kappa$ show dependence on the misorientation angle between the graphene and the h-BN. A quantitative result for this dependence is not available at present, but it is important to note that even a relatively small variation in $\kappa$ can lead to a large variation in the transmission term $e^{-2\kappa d}$. Thus, it is possible that the tunneling will be strongly confined to a narrow angular range of lateral wavevectors in the h-BN. To achieve those particular wavevectors in the h-BN, phonon scattering (or phonon-assisted tunneling) of the graphene states may play an important role. The presence of a linear background current in the measured characteristics, much greater than what is obtained in the simulated current as discussed at the end of Section IV, possibly provides evidence of such phonon participation in the tunneling process.

## Acknowledgements

This work was supported by STARnet, a Semiconductor Research Corporation program sponsored by MARCO and DARPA. Discussions with S. Satpathy on the theory of tunneling are gratefully acknowledged, and we thank P. Mende for a careful reading of the manuscript.

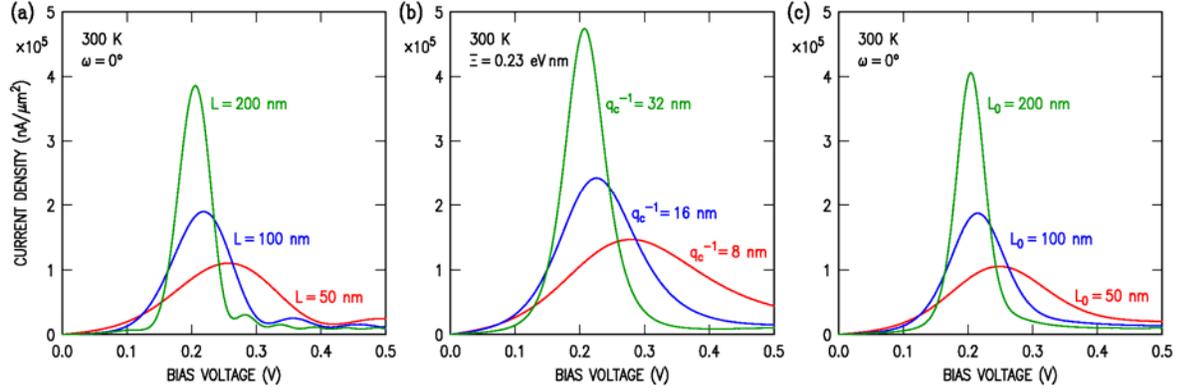

FIG. 1. Comparison of theoretical tunneling currents as a function of the bias voltage across two graphene electrodes separated by a h-BN insulator using the theories of (a) Ref. [3] with zero misorientation, varying coherence length $L$; (b) Ref. [6] with amplitude $\Xi = \hbar^2 \kappa / 2m$, varying $q_c = 2\pi / L$, and with the modified form of the scattering potential given in Eq. (10). In panel (c), we show the equivalence of the two theories by computing the total current of a device with a log-normal distribution of grain sizes with mean coherence length $L_0$ and variance $L_0^2 / 10$ with the current for each grain size computed using the theory of Ref. [3].

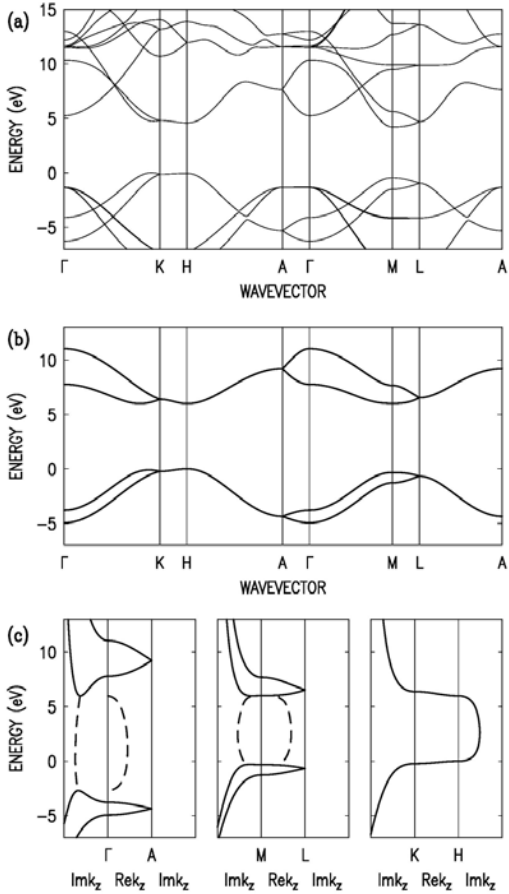

FIG. 2. (a) Band structure of hexagonal boron nitride, computed with density-functional theory. (b) Band structure from a tight-binding model, including only $p_z$ basis states. (c) Complex band structure from the tight-binding model, along the $\Gamma$A, ML, and KH directions. The right-hand and left-hand panels for each direction show the band structure with varying imaginary part of $k_z$. In these panels, solid lines denote bands for which $\mathrm{Re}(k_z)$ is constant, equal to the value at the point where the right- or left-hand panel joins the center panel. Dashed lines indicate bands for which both $\mathrm{Im}(k_z)$ and $\mathrm{Re}(k_z)$ is varying, in accordance to the lines in the respective $\mathrm{Im}(k_z)$ and $\mathrm{Re}(k_z)$ panels. In the each panel where $\mathrm{Im}(k_z)$ is varying, the range plotted is twice as large as in the corresponding panel showing $\mathrm{Re}(k_z)$

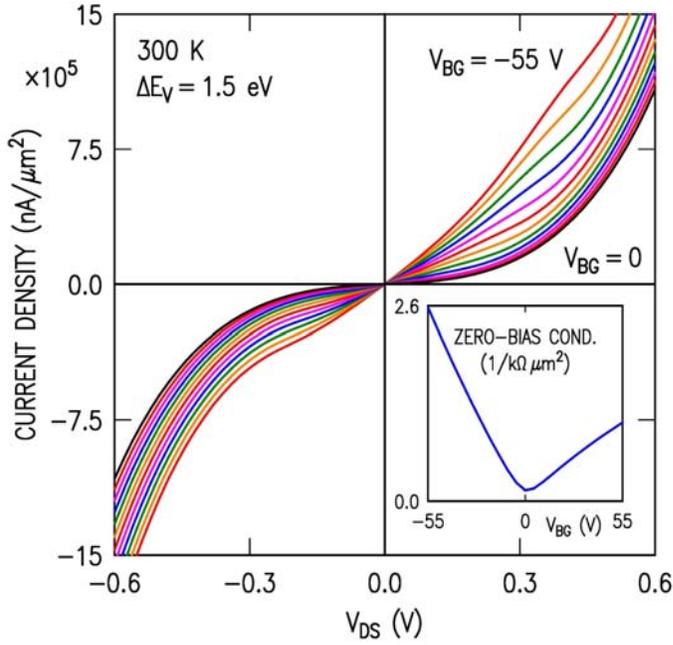

FIG. 3. Theoretical simulation of a GIG device with a back gate, corresponding to Fig. 4 of Ref. [28]. The simulated structure consists of the top layer of graphene, 4 layers of h-BN, the bottom layer of graphene, and 20 nm of h-BN on a silicon substrate (back gate) with a 300 nm $SiO_2$ dielectric film. Both graphene layers are assumed to be undoped. Curves are shown for $V_{BG} = -55$ to 0 V in 5 V increments. Zero-bias conductance vs. gate voltage is shown in the inset. The valence band offset that best fits with the experimental data is found to be $\Delta E_V = 1.5$ eV.

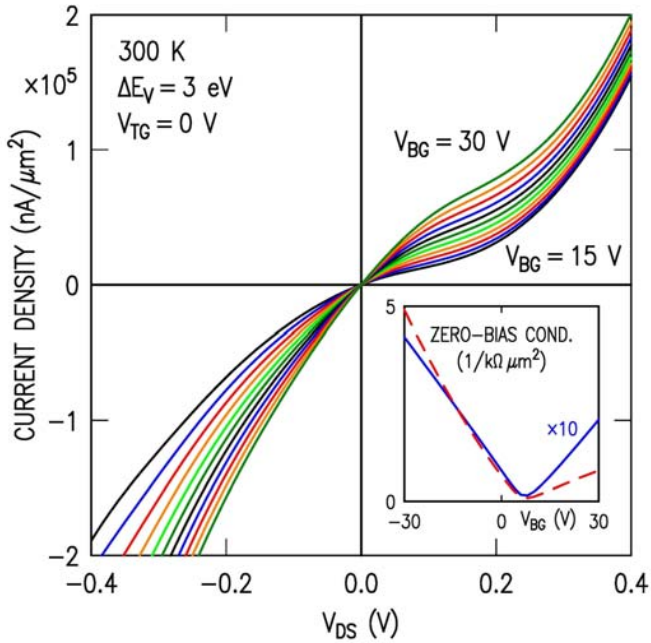

FIG. 4. Theoretical simulation of a GIG device with top and bottom gates (denoted $V_{TG}$ and $V_{BG}$), corresponding to Fig. 5 of Ref. [26]. The simulated structure consists of a top gate, 10 nm of a $HfO_2$ gate dielectric, the top layer of graphene, 4 layers of h-BN, and the bottom layer of graphene on a doped silicon substrate (back gate) with a 90 nm $SiO_2$ dielectric film. Both graphene layers are $p$-type doped with a carrier density of $p = 7.4 \times 10^{11}$ cm$^{-2}$ in each layer. Curves are shown for $V_{BG} = 15$ to 30 V in 1.5 V increments. Zero-bias conductance vs. back gate voltage is shown in the inset for a valence band offset of 3 eV (solid) and 1.5 eV (dashed). The best agreement with experiment (shown in main figure) is obtained for a valence band offset at midgap, $\Delta E_V = 3$ eV.

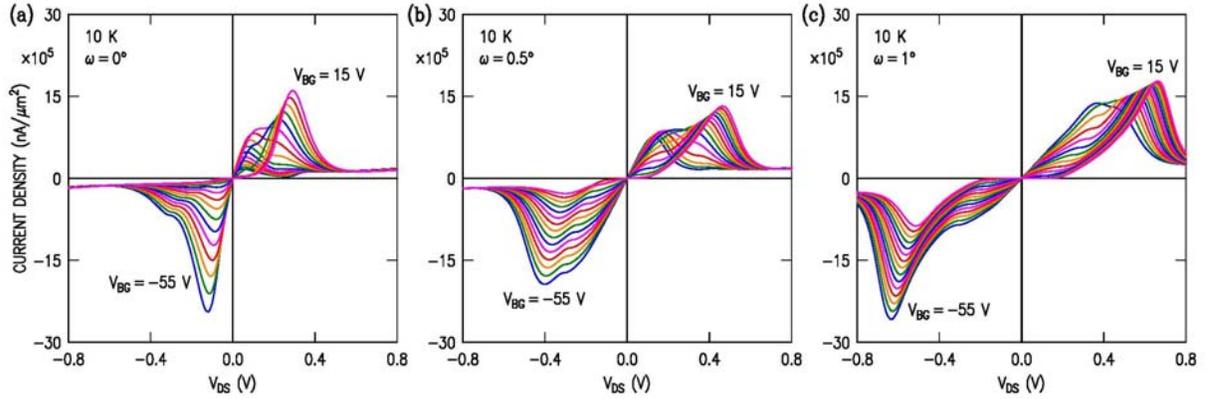

FIG. 5. Simulation of a gated GIG device corresponding to Fig. 1 of Ref. [6] that exhibits NDR. Tunneling characteristics are shown for (a) zero misorientation, (b) 0.5° of misorientation, and (c) 1.0° of misorientation between the graphene sheets. Curves are shown for $V_{BG} = -55$ to +15 V in 5 V increments. Computations are performed at low temperature to match with experiment. The device structure is identical to that of Fig. 3, but with doping in the top and bottom layers of graphene set to $p = 1.0 \times 10^{12}$ cm$^{-2}$ and $n = 4.4 \times 10^{11}$ cm$^{-2}$, respectively, as in Ref. [6].